# Rare-earth solid-state qubits

S. Bertaina[1,2], S. Gambarelli[3], A. Tkachuk[4], **I.N.** Kurkin[5], B. Malkin[5], A. Stepanov[6], and B. Barbara[1].

[1] Institut Néel, Département Nanosciences, CNRS, 25 Ave. des Martyrs, BP166, 38042 Grenoble Cedex 9, France.

[2] Ecole Nationale Supérieure de Physique, INP-Grenoble, Minatec 3 parvis Louis Néel, BP 257, 38016 Grenoble Cedex 9, France.

[3] Laboratoire de Chimie Inorganique et Biologie (UMR-E 3 CEA-UJF), DRFMC, CEA-Grenoble, 17 rue des Martyrs 38054 Grenoble cedex 9, France.

[4] S.I. Vavilov State Optical Institute, St.Petersburg 199034, Russian Federation.

[5] Kazan State University, Kazan 420008, Russian Federation.

[6] Laboratoire de Matériaux et Microélectronique de Provence, Faculté St Jérôme, C142, 13397, Marseille Cedex 20, France.

**Quantum bits (qubits) are the basic building blocks of any quantum computer. Superconducting qubits have been created with a 'top-down' approach that integrates superconducting devices into macroscopic electrical circuits [1-3], whereas electron-spin qubits have been demonstrated in quantum dots [4-6]. The phase coherence time ($\tau_2$) and the single qubit figure of merit ($Q_M$) of superconducting and electron-spin qubits are similar --- $\tau_2 \sim \mu$s and $Q_M \sim$10-1000 below 100mK --- and it should be possible to scale-up these systems, which is essential for the development of any useful quantum computer. Bottom-up approaches based on dilute ensembles of spins have achieved much larger values of $\tau_2$ (up to tens of ms) [7, 8], but these systems cannot be scaled up, although some proposals for qubits based on 2D nanostructures should be scalable [9-11]. Here we report that a new family of spin qubits based on rare-earth ions demonstrates values of $\tau_2$ (~ 50$\mu$s) and $Q_M$ (~1400) at 2.5 K, which suggests that rare-earth qubits may, in principle, be suitable for scalable quantum information processing at $^4$He temperatures.**

In general, a spin qubit state is a linear superposition of the two spin states of an electron $|\uparrow>$ and $|\downarrow>$. This means that the qubit can be represented as $|\psi_s> = \alpha|\uparrow> + \beta|\downarrow>$, where $\alpha$ and $\beta$ are probability amplitudes, and $|\alpha|^2 + |\beta|^2 = 1$. When measuring this qubit, the probability of outcome $|\uparrow>$ (or $|\downarrow>$) is $|\alpha|^2$ (or $|\beta|^2$). In rare-earth (RE) systems the total spin, **S**,

is no longer a good quantum number because the spin-orbit coupling between **S** and the total orbital angular momentum, **L**, is larger than the coupling of **L** with the electric field gradient of environmental ionic charges (crystal-field). The good quantum number is the total angular momentum, **J**=**L**+**S**, which is coupled with the crystal-field through **L**. The RE qubit states are therefore crystal-field states. In addition, RE often show isotopes with a nuclear spin, **I,** which has large interactions with **J** (hyperfine interactions) leading to electro-nuclear crystal-field states with wave functions $|\Psi_{en}>$ (see the crystal-field background section). Qubits based on these electro-nuclear states differ from typical spin qubits in several ways: (i) the crystal-field strongly affects the Rabi frequencies that depend on the direction and the strength of applied magnetic fields and electric field gradients, and this could open up new possibilities for scaling; (ii) the hyperfine interactions produce up to $3(2I+1)-2$ qubits per RE, all with slightly different resonance frequencies, which means that it should be quite easy to selectively address them with superimposed (low) field pulses; (iii) due to their large magnetic moment (~$10\mu_B$), it should be simple to manipulate RE qubits; (iv) the single qubit figure of merit, $Q_M$, should be large enough to allow quantum computer processing at $^4$He temperatures ($Q_M$ is the number of coherent single-qubit operations, defined as $\Omega_R\tau_2/\pi$, where $\Omega_R$ is the Rabi frequency: equivalently, it is the coherence time divided by half the Rabi period).

This work is an extension of a previous research that explored the quantum tunnelling of the magnetization in $Mn_{12}$-ac and Ho:YLiF$_4$ [12-14]. Due to the strong hyperfine interactions in the latter system, **J** tunnels simultaneously with **I** (electro-nuclear tunneling). The system chosen to illustrate the concept of RE qubits consists in Er$^{3+}$ ions (J = 15/2 and $g_J$ = 6/5) diluted in a single crystalline matrix of CaWO$_4$, which is isomorphic with YLiF$_4$. The main reason for replacing YLiF$_4$ with CaWO$_4$ is to reduce the proportion of nuclear spins, which are an important source of decoherence [15] (the phenomenon by which a quantum system seems to be classical as a result of interactions with its environment ).

Continuous wave electron paramagnetic resonance (CW-EPR) measurements were first performed in Er$^{3+}$:CaWO$_4$. The transitions for the isotopes with I=0 and I=7/2 were observed at $^4$He temperatures using a Bruker X-band spectrometer at 9.7 GHz. These transitions occur either between pure crystal-field levels (I=0) or between electro-nuclear crystal-field sublevels (I=7/2) (see background section). In both cases the observed line-width is small enough for the lifetime of the levels to be much larger than calculated periods of Rabi oscillations (weak decoherence).

In order to observe these oscillations, a series of experiments were performed in pulsed wave EPR (PW-EPR) mode. Eight transitions were observed (Fig 1).

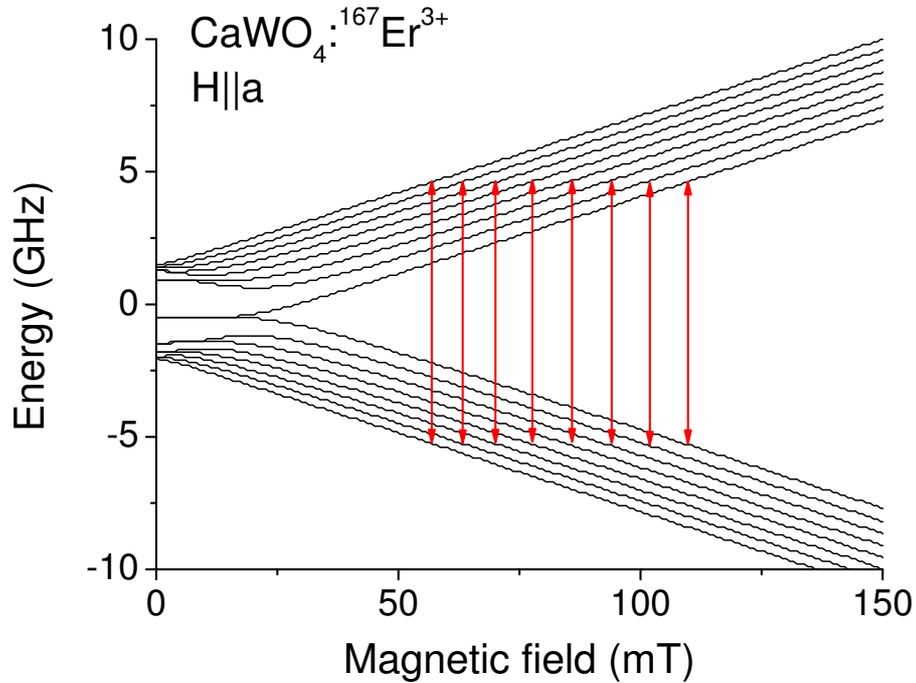

**Fig. 1 Energy levels and Rabi frequencies for the erbium-doped rare-earth system $^{167}Er^{3+}$:CaWO$_4$. a,** Energy spectrum calculated for a magnetic field perpendicular to the c-axis. In zero field the spectrum contains 16 electro-nuclear states ((2S+1)(2I+1) with S=1/2 and I=7/2) consisting in a singlet, 7 doublets and another singlet (nine sublevels). The fourth doublet, near the centre of the figure, is well separated from the other levels. When the Zeeman splitting caused by the magnetic field becomes larger than the hyperfine splitting, which sets the energy scale at zero-field, the levels vary linearly with the magnetic field, which gives 8 states with effective spin ½ and 8 states with effective spin -1/2: each of these states is labelled by the nuclear spin projection, $m_I$, which increases from -7/2 for the two states at the centre of the figure to +7/2 for the lower- and upper-most states. EPR transitions between spins ±1/2 and $\Delta m_I =0$ are represented by the vertical arrows. **b,** Rabi frequencies, measured vs static field H // a and ac-field $\mu_0 h$ =0.12 mT // b, on a single-crystal of $Er^{3+}$:CaWO$_4$ (2x2.5x3mm$^3$, 10$^{-5}$ atomic % Er). They show an intense central peak (for the isotopes I=0) and 8 smaller peaks separated by $\Delta H$ ~ 6-8 mT (for the isotope I=7/2, $^{167}Er^{3+}$). Exact diagonalization of (1) (see background section), permits accurate calculation of these frequencies (using the crystal-field and hyperfine constants only [30,32]); one finds $\Omega_R/2\pi$ = 17.546, 17.302, 17.166, 17.115, 17.137, 17.238, 17.394, and 17.605 MHz. The colour scale shows the proportion of ions with Rabi frequency $\Omega_R$ at a given magnetic field (white < 80, blue = 80, red> 800 arbitrary units).

An example of the measured Rabi oscillations [16] is given in Fig. 2, for I=0, where the z component of the magnetization, $M_z$, is plotted vs time. It is possible to fit the data to

$$\langle M_z \rangle = M_{z(t=0)}\, e^{-t/\tau_R} \sin(\Omega_R t) \qquad (1)$$

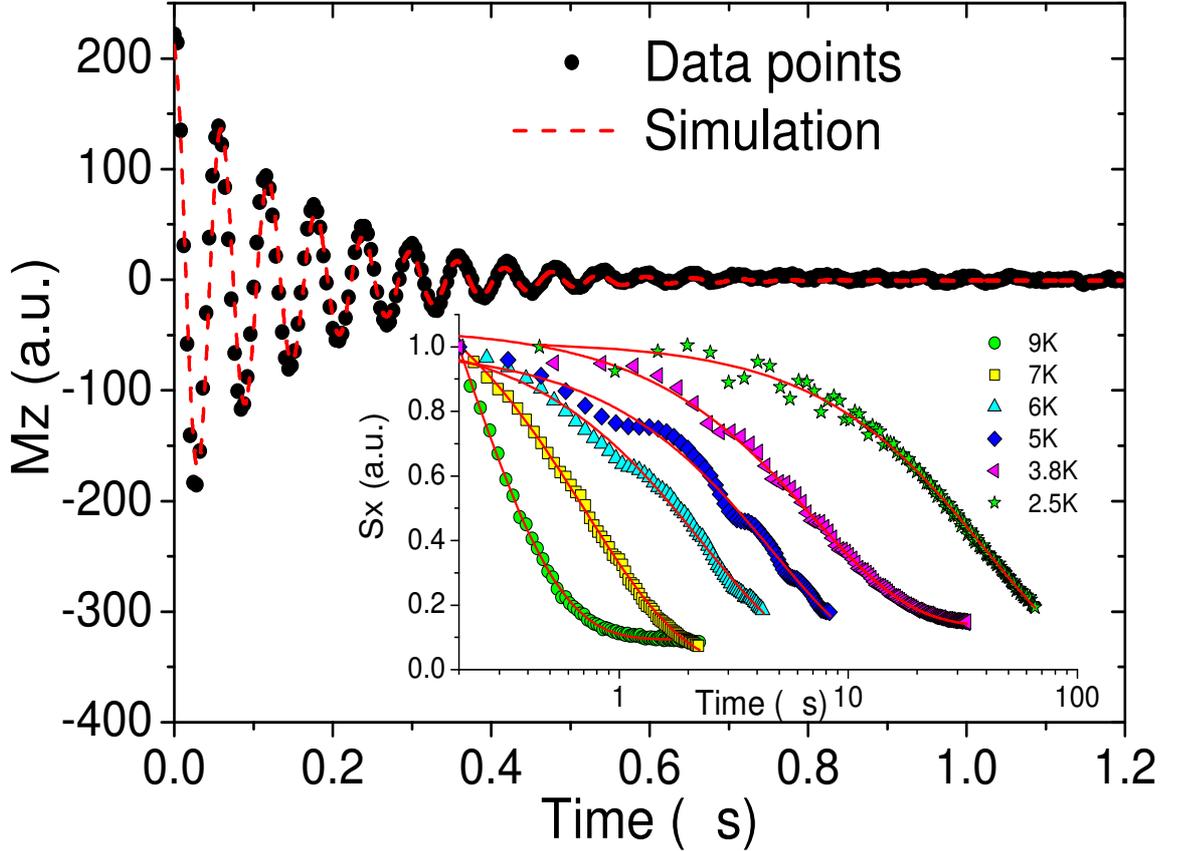

**Fig 2 Rabi oscillations and coherence times**. Rabi oscillations measured on I=0 isotopes of the same single-crystal for $\mu_0 H$=0.522 T //c, $\mu_0 h$ =0.15 mT //b and T=3.5 K. These oscillations are obtained by the application of a nutation pulse of length t followed, after a delay greater than $\tau_2$ (permitting the transverses spin components to relax), by a $\pi/2$-$\pi$ sequence. The resulting echo intensity is averaged over ~$10^3$ measurements, giving the z-component of the nutating magnetization at time t ($M_z$). The dashed line is a fit to equ. (1) (see text) giving an exponential decay time $\tau_R$= 0.2 µs << $\tau_2$ ~ 7µs (see Fig. 3). **Inset**: Decay of the transverse spin component, $S_x$, obtained by a conventional spin-echo method at different temperatures, showing that the coherence time $\tau_2$ reaches the 100µs scale at $^4$He temperatures. Weak superimposed oscillations come from the ESEEM effect (Electron Spin Echo Envelope Modulation) [34] produced by the super-hyperfine coupling with second neighbour W nuclear spins. One can verify that the oscillation frequency perfectly matches W nucleus spin Larmor frequency in the applied field (small super hyperfine limit).

using a single exponential damping parameter $\tau_R \sim 0.2$ μs ($\Omega_R$ having been previously obtained from a Fourier transform of the data). Other experiments performed at different microwave powers show that $\tau_R$ increases as the power decreases, while the number of Rabi oscillations, N(c) (where c is the Er concentration) remains nearly unchanged i.e. $N(c) \sim \tau_R(c)\Omega_R$ with $N(c) \sim 20$ in the example of Fig. 2. This increase of $\tau_R$ is always limited by $\tau_2$ (Fig. 3). All of this suggests the phenomenological expression

$$1/\tau_R(c) \sim \Omega_R/N(c) + 1/\tau_2(c) \qquad (2)$$

where $\tau_R(c)$, N(c), and $\tau_2(c)$ are concentration-dependent. Rabi oscillations are lost for $t \gg \tau_2$ in the low power limit where $\Omega_R \to 0$, and for $t \gg N(c)/\Omega_R$ in the large power limit where $\Omega_R \gg N(c)/\tau_2$. In the first case $\tau_2$ should be limited by RE spin-diffusion due to long range dipolar interactions, as in NMR. In the second case, the observed behaviour is characteristic of inhomogeneous nutation frequency. In fact, a weak random crystal-field, responsible for the CW line-width [17,18] feeds into some distribution of the $|J,m_J,I,m_I\rangle$ coefficients resulting in destructive interference of Rabi oscillations ($\Omega_R \propto \langle \phi_1, m_I | J_+ | \phi_2, m_I' \rangle$, see background section) which go out of phase after a certain number of periods. However, the number of oscillations N(c) depends on concentration, indicating that dipolar interactions must also be taken into account.

Recently, a model relying on the assumption that each spin experiences a stochastic field of mean-square amplitude β, oscillating at the resonance frequency ω, led to an expression,

$$1/\tau_R = \beta\Omega_R + 1/2\tau_2 \qquad (3)$$

very similar to (2) [19]. This linear dependence on $\Omega_R$ was tested on pure $S=1/2$ spins in amorphous-$SiO_2$ containing E' centres where a concentration effect has also been obtained [20]. In the frame of the present study, the origin of the stochastic field should be related to both crystal-field distribution and dipolar interactions [21]. In order to check (2) and (3) more carefully, $1/\tau_R$ vs $\Omega_R$ is plotted for two different directions of the microwave field h (Fig. 3, inset; see also Fig. 4). The obtained curve is continuous, showing that the damping rate scales with the Rabi frequency (and not with the microwave field h when dipole matrix elements are different) according to an S-shaped curve of e.g. the type $1/\tau_R = 1/\tau_2 f(\Omega_R \tau_2)$, with a progressive saturation at $\tau_2$ when $\Omega_R \to 0$. The dependence of $\Omega_R$ with the direction of the microwave field, h, is demonstrated in Fig. 4 where a simple rotation from h // b to h // c reduces the Rabi frequency by the factor $\Omega_{R//b}/\Omega_{R//c} \sim 6$. This ratio is slightly smaller than the one derived from the proportionality $\Omega_{R//b}/\Omega_{R//c} \sim g_{effb}/g_{effc} \sim 6.7$ because $\Omega_{R//b}$ drops in a few degrees from its maximum value $\Omega_{R//b} \sim 6.7\Omega_{R//c}$ to its minimum value $\sim \Omega_{R//c}$. A better agreement would simply require better angular accuracy in the crystal orientation.

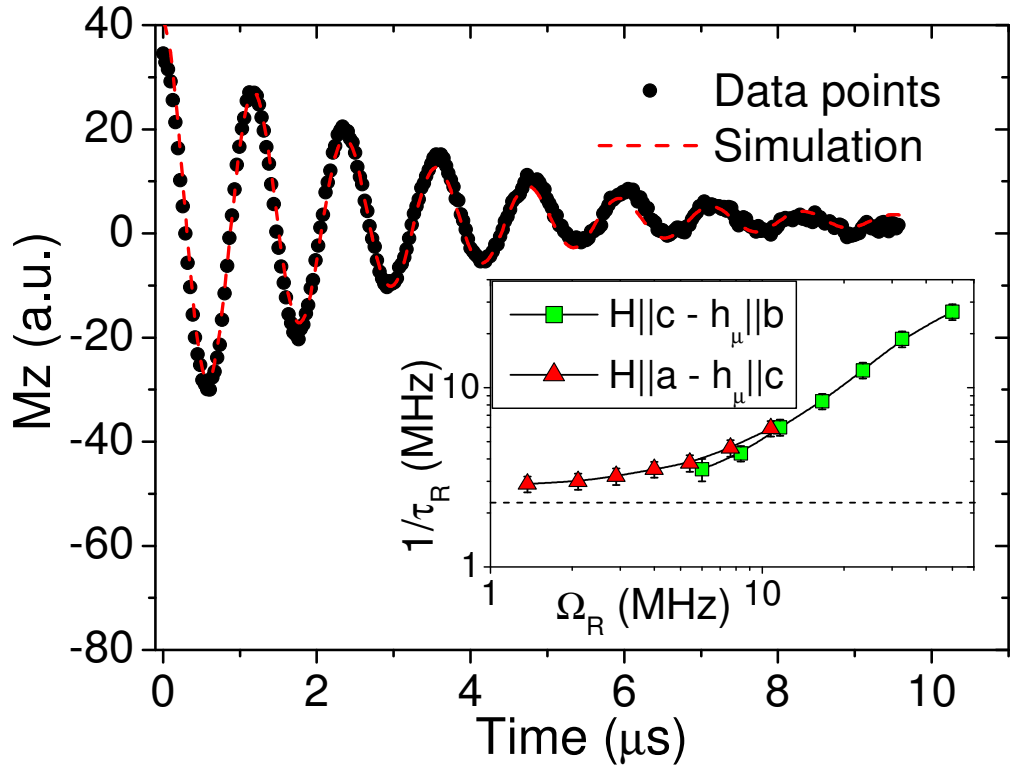

**Fig. 3 Changing the damping time with the microwave power.** When the experiment in figure 2 is repeated with the microwave field reduced by a factor of 20, the period of the Rabi oscillations becomes longer (by the same factor of 20), but the number of oscillations remains of the order of 20 (up to 20 μs). The same fit as in Fig. 2 gives $\tau_R \sim 3$ μs, which is comparable with the $\tau_2 \sim 7$μs obtained in spin-echo measurements under the same experimental conditions. **Inset**: Damping rate of). the Rabi oscillations, $1/\tau_R$, plotted against the Rabi frequency, $\Omega_R$, for two directions of the microwave field. The continuity of the curve proves that $1/\tau_R$ depends on $\Omega_R$ only and tends to $1/\tau_2$ at low microwave power (dashed line)

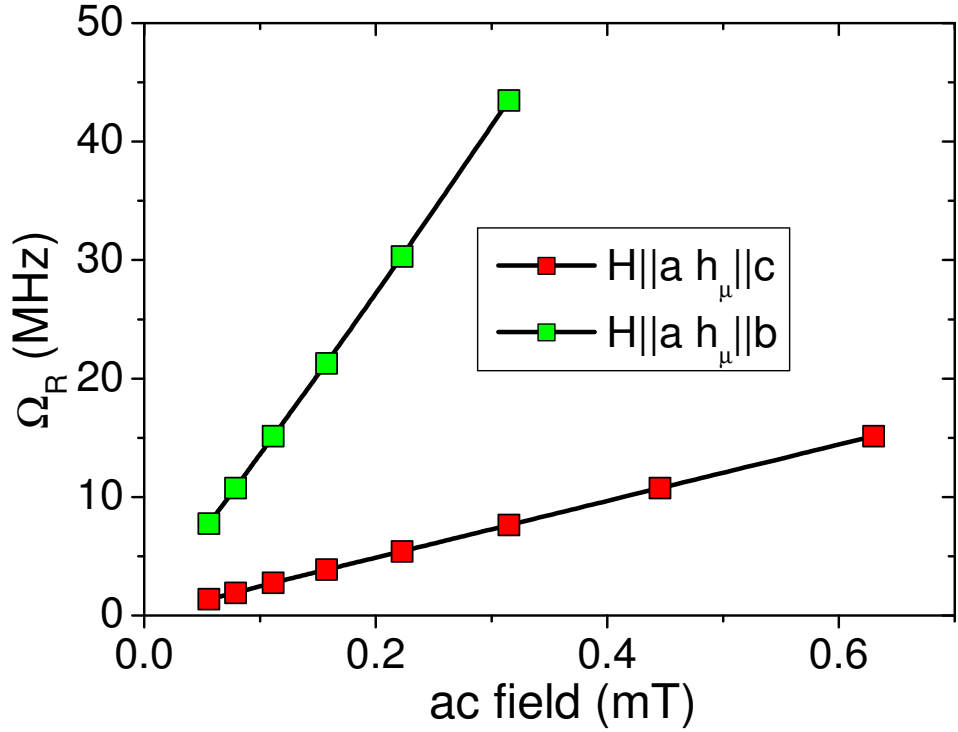

**Fig. 4 Maximum and minimum coupling of the microwave field to Er effective spins and direction-dependant Rabi frequencies.** Rabi frequency measured for two directions of the microwave field, $h_\mu$, on a single-crystal with atomic Er concentration of $5.10^{-4}$. The ac field was calibrated by comparison with a coal sample. Due to the "easy" plane anisotropy (see background section), the coupling between Er effective spins and the microwave field is maximum when the latter is in the easy-plane (giving large $\Omega_R$) and minimum when it is perpendicular to it (giving small $\Omega_R$).

Finally, RE qubits have large $Q_M$ at $^4$He temperatures and in principle they should be scalable. Indeed $\tau_2$ increases with dilution and cooling (Fig.2b); an extrapolation down to 1.5 K for a concentration of $10^{-6}$ atomic Er:CaWO$_4$ gives $Q_M \sim 10^4$ which is enough for quantum information processing. Moreover, RE qubits could in principle be selectively addressed and their couplings manipulated, according to variants of existing proposals and realizations (10,11,4-6]. As a matter of fact, they could be inserted in all kinds of matrices structured by lithography, including films, quantum dots or nanowires of semi-conducting Si [22] or GaN [23], and coupled by controlled carrier injection through gate voltage [24]. They could be addressed selectively by application of: (i) local field pulses of amplitude < 25 mT adding algebraically to the static field (this is limited to n ≤ 3(2I+1)-2 qubits, Fig. 1); (ii) continuous electric field gradients for n>3(2I+1)-2. A gradient of 10 mV/(nm)$^2$ is enough to modify the crystal-field parameters by ~10% in most matrices and therefore the resonance frequency. Interestingly, the 3(2I+1)-2 Rabi oscillations of each $^{167}$Er (Fig. 1) may also be used to implement Grover's algorithm [25] on single RE ions (this is a general property of electro-nuclear RE qubits with I≠0). Spin-state detection could follow schemes like those in [4,6] but alternative ways using fast photo-luminescent properties of RE [22,23,26] might ultimately be better. Finally, instead of dots one might also use single molecules containing a RE ion [27].

In conclusion, Rabi oscillations of the angular moment J=15/2 of Er:CaWO$_4$ have been observed for the first time and analysed, evincing a new type of anisotropic electro-nuclear spin qubits. Isotopes with I=0 give a single purely electronic Rabi frequency (single qubit, $\Delta M_J = \pm 1$), while the isotope I=7/2 ($^{167}$Er) gives a set of eight electro-nuclear frequencies (eight qubits, $\Delta M_I = \pm 1$ and $\Delta M_I = 0$) which are addressed independently. Because the spin-orbit coupling, the magnetic moments and the hyperfine interactions are all large, it should be possible to couple and address selectively a large number of RE qubits using weak electric and magnetic fields. Furthermore, each RE ion could be used to implement Grover's algorithm. All this, together with large $Q_M$ factors (~ 10$^3$ -10$^4$ between 2.5-1.5 K), suggests that RE qubits are good candidates for implementation of quantum computation at $^4$He temperatures.

**Crystal-field background**

1. Hamiltonian

The single-ion Hamiltonian for Er$^{3+}$:CaWO$_4$ (tetragonal space group I4$_1$/a and S$_4$ point symmetry [28]) contains crystal-field, hyperfine and Zeeman terms:

$$H_{CF} = \alpha_J B_2^0 O_2^0 + \beta_J (B_4^0 O_4^0 + B_4^4 O_4^4) + \gamma_J (B_6^0 O_6^0 + B_6^4 O_6^4 + B_6^{-4} O_6^{-4}) + A_J \mathbf{I} \cdot \mathbf{J} + g_J \mu_B \mu_0 \mathbf{J} \mathbf{H} . \quad (1)$$

The $O_l^m$ are the Stevens' equivalent operators with the reduced matrix elements $\alpha_J$, $\beta_J$, $\gamma_J$ [29], and the $B_l^m$ are the crystal-field parameters determined by high resolution optical spectroscopy

($B_2^0$ =231 cm$^{-1}$, $B_4^0$ = -90 cm$^{-1}$, $B_4^4$ = ± 852 cm$^{-1}$, $B_6^0$ = -0.6 cm$^{-1}$, $B_6^4$ = ± 396 cm$^{-1}$, and $B_6^{-4}$ = ± 75 cm$^{-1}$ [30]).

2. Energy spectra and wave functions

Exact diagonalization of the 16x16 matrix of (1) with **I**=**H**=0, shows an easy plane perpendicular to the c-axis with a doublet ground-state of wave functions $|\phi_1\rangle$ and $|\phi_2\rangle$. This doublet, with effective spin ½ and anisotropic $g_{eff}$ tensor ($g_{//}$ = 1.247, $g_\perp$ = 8.38 [31]), permits a single EPR transition ($\Delta m_J = \pm 1$) which can be observed on I=0 isotopes (~ 77%). The Rabi frequency is given by $\Omega_R = 2g_J \mu_B \langle \phi_1 | J_\mu h_\mu | \phi_2 \rangle / \hbar \propto g_{eff}$, where $\hbar$ is the Planck's constant. Natural Er also contains $^{167}$Er with I=7/2 (~23%) and $A_J$ = - 4.16 10$^{-3}$ cm$^{-1}$ (-125 MHz) [32]. In this case the 128x128 matrix leads to the energy spectrum of Fig 1a. The degeneracy is completely removed by **H** and the new set of wave functions $|\Psi_{en}\rangle = \Sigma b_i |J, m_J, I, m_I\rangle$ on the space product $|L,S,J,m_J\rangle \otimes |I,m_I\rangle$ differs from $|\phi_1\rangle$ and $|\phi_2\rangle$ owing to the nuclear degrees of freedom. Fig. 1 also shows that 3(2I+1)-2 EPR transitions are allowed, giving for I=7/2, 8 transitions with conservation of I ($\Delta m_J = \pm 1$ and $\Delta m_I = 0$) and 14 transitions without ($\Delta m_J = \pm 1$ and $\Delta m_I = \pm 1$).

**Acknowledgements**

The authors acknowledge the support of INTAS contract N° 2003/03-51-4943. B.M. and I.K. acknowledge the Ministry of Education and Science of the Russian Federation (project RNP 2.1.1.7348) and B.B. the interdisciplinary European Network of Excellence "MAGMANet" for support during the first year.
A.T. provided the samples. S.B. and S.G. performed the experiments, analysed and discussed them with A.S., I.K., B.M. and BB. B.B. proposed this study and the manuscript which was commented on by all the authors.